\pdfoutput=1

\documentclass[10pt, conference]{IEEEtran}
\ifCLASSINFOpdf
\else
\fi

\usepackage{tikz}

\usepackage{graphicx}
\usepackage{balance}  
\usepackage[caption=false,font=footnotesize]{subfig}

\usepackage[noend]{algpseudocode}
\usepackage{algorithm}

\usepackage{multirow}
\usepackage{amsfonts}

\begin{document}
%
\title{HYPE: Massive Hypergraph Partitioning  with Neighborhood Expansion}



\author{\IEEEauthorblockN{Christian Mayer, Ruben Mayer*, Sukanya Bhowmik, Lukas Epple, and Kurt Rothermel}
	\IEEEauthorblockA{Institute of Parallel and Distributed Systems, University of Stuttgart, Germany\\
		email: \{firstname.lastname\}@ipvs.uni-stuttgart.de\\ 
		\space * Technical University of Munich, Germany\\email: ruben.mayer@tum.de}}


%


\maketitle

\begin{tikzpicture}
\begin{scope}[overlay]
\footnotesize
\node[text width=40cm] at ([yshift=-19.0cm,xshift=9cm]current page.south) {\copyright 2018 IEEE. Personal use of this material is permitted. Permission from IEEE must be obtained for all other uses,  in any current or future media, \newline including reprinting/republishing this material for advertising or promotional purposes,  creating new collective works, for resale or redistribution \newline to servers or lists, or reuse of any copyrighted component of this work in other works. This is the authors' version of the work. \newline The definite version is published in Proceedings of 2018 IEEE International Conference on Big Data (BigData '18).};
\end{scope}
\end{tikzpicture}

\begin{abstract}
Many important real-world applications---such as social networks or distributed data bases---can be modeled as hypergraphs. In such a model, vertices represent entities---such as users or data records---whereas hyperedges model a group membership of the vertices---such as the authorship in a specific topic or the membership of a data record in a specific replicated shard. To optimize such applications, we need an efficient and effective solution to the NP-hard balanced k-way hypergraph partitioning problem. However, existing hypergraph partitioners that scale to very large graphs do not effectively exploit the hypergraph structure when performing the partitioning decisions. We propose HYPE, a hypergraph partitionier that exploits the neighborhood relations between vertices in the hypergraph using an efficient implementation of neighborhood expansion. HYPE improves partitioning quality by up to $95\%$ and reduces runtime by up to $39\%$ compared to streaming partitioning. 
\end{abstract}


%
\IEEEpeerreviewmaketitle


\section{Introduction}
\label{sec:introduction}

Many real-world applications model problems as hypergraphs where connections between vertices are multi-dimen-sional, i.e., each vertex can directly communicate to n vertices in a group (called ``hyperedge'') and each vertex can be in multiple groups. Hypergraph partitioning~\cite{Karypis:1999:MKW:309847.309954, vastenhouw2005two, TRIFUNOVIC2008563, 780863, 1639359, hs2017sea, kabiljo2017social, Alistarh:2015:SMH:2969442.2969452} deals with the problem of optimally dividing a hypergraph into a set of equally-sized components. Applications of hypergraph partitioning arise in diverse areas such as VLSI design placement~\cite{karypis1999multilevel}, optimizing the task and data placement in workflows~\cite{Catalyurek:2011:IDP:1996014.1996022}, minimizing the number of transactions in distributed data placement~\cite{Curino:2010:SWA:1920841.1920853}, optimizing storage sharding in distributed data bases~\cite{kabiljo2017social}, and as a necessary preprocessing step in distributed hypergraph processing~\cite{7373388}.

Formally, dividing the hypergraph is denoted as the \textit{balanced k-way hypergraph partitioning problem}. The goal is to divide the hypergraph into $k$ equally-sized partitions such that the number of times neighboring vertices are assigned to different partitions is minimal (this is denoted as the (k-1) metric, as introduced later).
The balanced k-way hypergraph partitioning problem is NP-hard. Hence, a heuristic approach is needed to solve the problem for massive hypergraphs.

In literature, a couple of heuristic hypergraph partitioning algorithms have been proposed, but they have shortcomings. Streaming hypergraph partitioning~\cite{Alistarh:2015:SMH:2969442.2969452} considers one vertex at a time from a stream of hypergraph vertices. Based on a scoring function, it greedily assigns each vertex from the stream to the partition that yields the best scoring. While this algorithm has low run-time, it does not consider all relationships between all vertices when deciding on the partitioning, so that partitioning quality suffers. 
A recent hypergraph partitioning algorithm, Social Hash Partitioner~\cite{kabiljo2017social}, considers the complete hypergraph at once. It iteratively performs random permutations of the current partitioning followed by a greedy optimization to choose a better permutation over a worse one. While this approach generally converges to some form of an improved partitioning and is highly scalable, we argue that random permutations may not be the most effective choice for the partitioning heuristic. 

In the related field of \emph{graph} partitioning, a recently proposed algorithm uses \emph{neighborhood expansion} to exploit structural properties of natural graphs~\cite{Zhang:2017:GEP:3097983.3098033}. Graphs can be regarded as special cases of hypergraphs, where each hyperedge contains only a single vertex. However, the original neighborhood expansion algorithm for graphs cannot be directly applied to hypergraphs. As hyperedges may contain a very large number of vertices, the neighborhood of a single vertex can be huge, rendering neighborhood expansion infeasible. 

In this paper, we are the first researchers who successfully apply neighborhood expansion to hypergraph partitioning. We propose HYPE, a hypergraph partitioning algorithm specifically tailored to real-world hypergraphs with skewed degree distribution. HYPE grows $k$ partitions based on the neighborhood relations in the hypergraph. We evaluate the performance of HYPE on a set of real-world hypergraphs, including a novel hypergraph data set consisting of authors and subreddits from the online board Reddit. Reddit is, to the best of our knowledge, the largest \emph{real-world hypergraph} that has been considered in evaluating hypergraph partitioning algorithms up to now.  
In our evaluations, we show that HYPE can partition very large hypergraphs efficiently with high quality. HYPE is 39\% faster and yields 95\% better partitioning quality than \emph{streaming partitioning}~\cite{Alistarh:2015:SMH:2969442.2969452}. We released the source code of HYPE as an open source project: https://github.com/mayerrn/HYPE.

\section{Problem Formulation}
\label{sec:problemFormulation}

In this section, we formulate the hypergraph partitioning problem addressed in this paper. 

\textbf{Problem Formulation: }
The hypergraph is given as $G=(V,E)$ with the set of vertices $V$ and the set of hyperedges $E \subset 2^V$. Given vertex $v \in V$, we denote the set of adjacent vertices, i.e., the set of neighbors of $v$, as $N(v) \subseteq V$.
The goal is to partition the hypergraph into $k$ partitions $P=\{p_0, p_1, ..., p_{k-1}\}$ by assigning vertices to partitions.
The assignment function $A:V \rightarrow P$ defines for each vertex in $V$ to which partition it is assigned. We write $A(v)=p_i$ if vertex $v$ is assigned to partition $p_i$.
Each hyperedge spans between 1 and $k$ partitions.
The optimization objective is the $(k-1)$-cut that sums over each hyperedge the number of times it is assigned to more than one partition, i.e., $\sum_{e \in E} |\{p \in P | \exists v \in E: A(v)=p\}|-1$.
We require that the assignment of vertices to partitions is balanced in the number of vertices assigned to a single partition, i.e., $\forall p_0, p_1 \in P: |\{v \in V | A(v)=p_0\}| < \lambda |\{v \in V | A(v)=p_1\}|$ for a small balancing factor $\lambda \in \mathbb{R}$.
This problem is denoted as \textit{balanced k-way hypergraph partitioning problem} and it is NP-hard~\cite{Andreev2006}.

\begin{figure}	
	\includegraphics[width=0.45\textwidth]{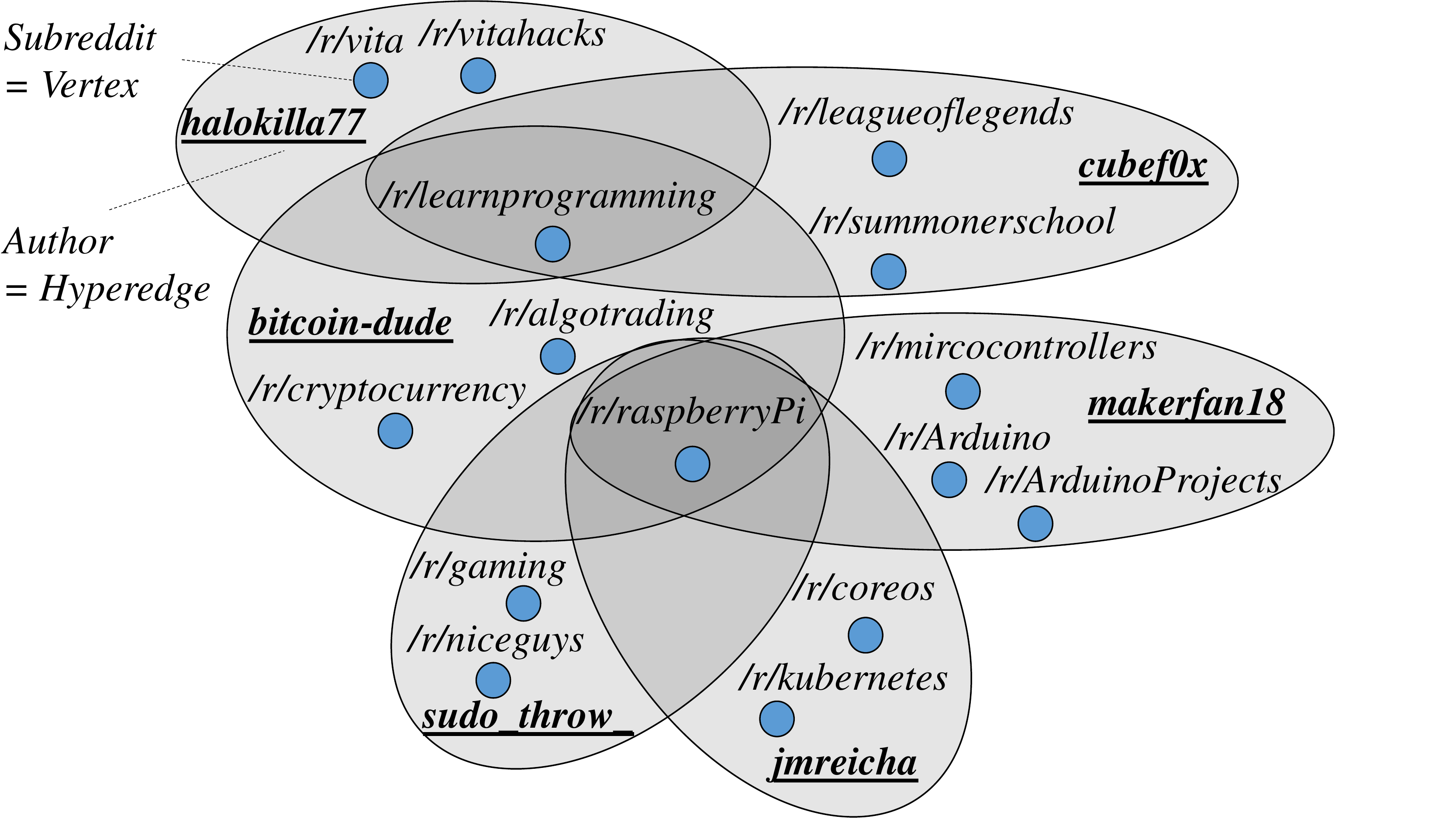}
	\caption{A small extract of the global Reddit graph.}
	\label{fig:redditExample}
	\vspace{-10pt}
\end{figure}

%

\textbf{Reddit Hypergraph Example:}
We give an example of the Reddit hypergraph in Figure~\ref{fig:redditExample}. The Reddit social network consists of a large number of subreddits where each subreddit concerns a certain topic. For example, in the subreddit \texttt{/r/learnprogramming}, authors write comments related to the topic of learning to program. Each author writes comments in an arbitrary number of subreddits and each subreddit is authored by an arbitrary number of users. 
One way to build a hypergraph out of the Reddit data set (see Section~\ref{sec:evaluations}) is to use subreddits as vertices and authors as hyperedges that connect these vertices.
This hypergraph representation provides valuable information about the similarity of subreddits. For instance, if many authors are active in two subreddits (e.g. \texttt{/r/Arduino} and \texttt{/r/ArduinoProjects}), i.e., many hyperedges overlap significantly in two vertices, it is likely that the two subreddits concern similar content.

The size of the hyperedges and the degree of the vertices resemble a power law degree distribution for real-world graphs such as Reddit, Stackoverflow, and Github.
Hence, most vertices have a small degree and most hyperedges have a small size. These parts of the graph with small degrees build relatively independent communities. In the reddit graph, there are local communities such as people who write in the \texttt{/r/Arduino} and \texttt{/r/ArduinoProjects} subreddits but not in, say the \texttt{/r/Baby} subreddit. This property of strong local communities is well-observed for graphs \cite{watts1998collective}---and it holds for real-world hypergraphs as well. Power law distributions are long-tailed, i.e., there are some hub vertices or edges with substantial sizes or degrees. In graph literature, it has been shown that focusing on optimal partitioning of the local communities at the expense of optimal placement of the hubs is a robust and effective partitioning strategy \cite{Gonzalez2012PowerGraph, Petroni:2015:HSP:2806416.2806424, Albert2000Error}. In Section~\ref{sec:algorithm}, we show how we exploit this observation in the HYPE partitioner.

\begin{table}
	\renewcommand{\arraystretch}{1.1}
	{
		\begin{center}
			\begin{tabular}{ | c | c | }
				\hline
				$G$ & Hypergraph $G=(V,E)$ \\ \hline
				$V$ & Set of vertices $V \subset \mathcal{N} \times \mathcal{N}$ \\ \hline
				$E$ & Set of hyperedges $E \subset 2^V$ \\ \hline
				$N(v)$ & Set of adjacent vertices of vertex $v \in V$ \\ \hline
				$N(X)$ & Union of all sets $N(x)$ for $x \in X$ \\ \hline
				$P$ & Set of partitions $p_i \in P$ with $|P|=k$ \\ \hline
				$A$ & Function assigning vertices $V$ to partitions $P$ \\ \hline
				$\lambda$ & Balancing factor \\ \hline
				$C$ & Current core set of vertices \\ \hline
				$F$ & Current fringe \\ \hline
				$d_{ext}(u,S)$ & External neighbors score of vertex $u$ \\ \hline
				$s$ & Maximal size of the fringe \\ \hline 
				$r$ & Number of fringe candidate vertices \\ \hline
			\end{tabular}
		\end{center}
	}
	\vspace{-5pt}
	\caption{Notation overview.}
	\label{tab:notation}
	\vspace{-25pt}
\end{table}
\section{The Algorithm}
\label{sec:algorithm}

In the following, we first explain the idea of neighborhood expansion in Section~\ref{sec:alg:ne}. We introduce our novel hypergraph partitioning algorithm HYPE in Section~\ref{sec:alg:hype} and discuss the balancing of hypergraph partitions in Section~\ref{sec:balancingConsiderations}. Finally, we present a more formal pseudocode notation of the HYPE algorithm in Section~\ref{sec:alg:pseudocode} and perform a complexity analysis in Section~\ref{sec:alg:complexity}.

\subsection{Neighborhood Expansion Idea}
\label{sec:alg:ne}

A practical method for high-quality \textit{graph} partitioning is \textit{neighborhood expansion} \cite{Zhang:2017:GEP:3097983.3098033}. 
The idea is to grow a core set of vertices via the neighborhood relation given by the graph structure. By exploiting the graph structure, the locality of vertices in the partition is maximized, i.e., neighboring vertices in the graph tend to reside on the same partition. The algorithm grows the core set one vertex at a time until the desired partition size is achieved. In order to partition the graph into $k$ partitions, the procedure of growing a core set $C_i$ is repeated $k$ times for $i \in \{0,1,...,k-1\}$.

We aim to adopt neighborhood expansion to hypergraph partitioning. To this end, we have to overcome a set of challenges which are related to the different structure of hypergraphs when compared to normal graphs. In particular, the number of neighbors of a vertex quickly explodes as the hyperedges contain multiple neighboring vertices at once. Before we explain in detail those challenges and our approach to tackle them, we sketch the basic idea of neighborhood expansion in the following.

\begin{figure}	
	\includegraphics[width=0.45\textwidth, trim= 0pt 130pt 0pt 0pt, clip=true]{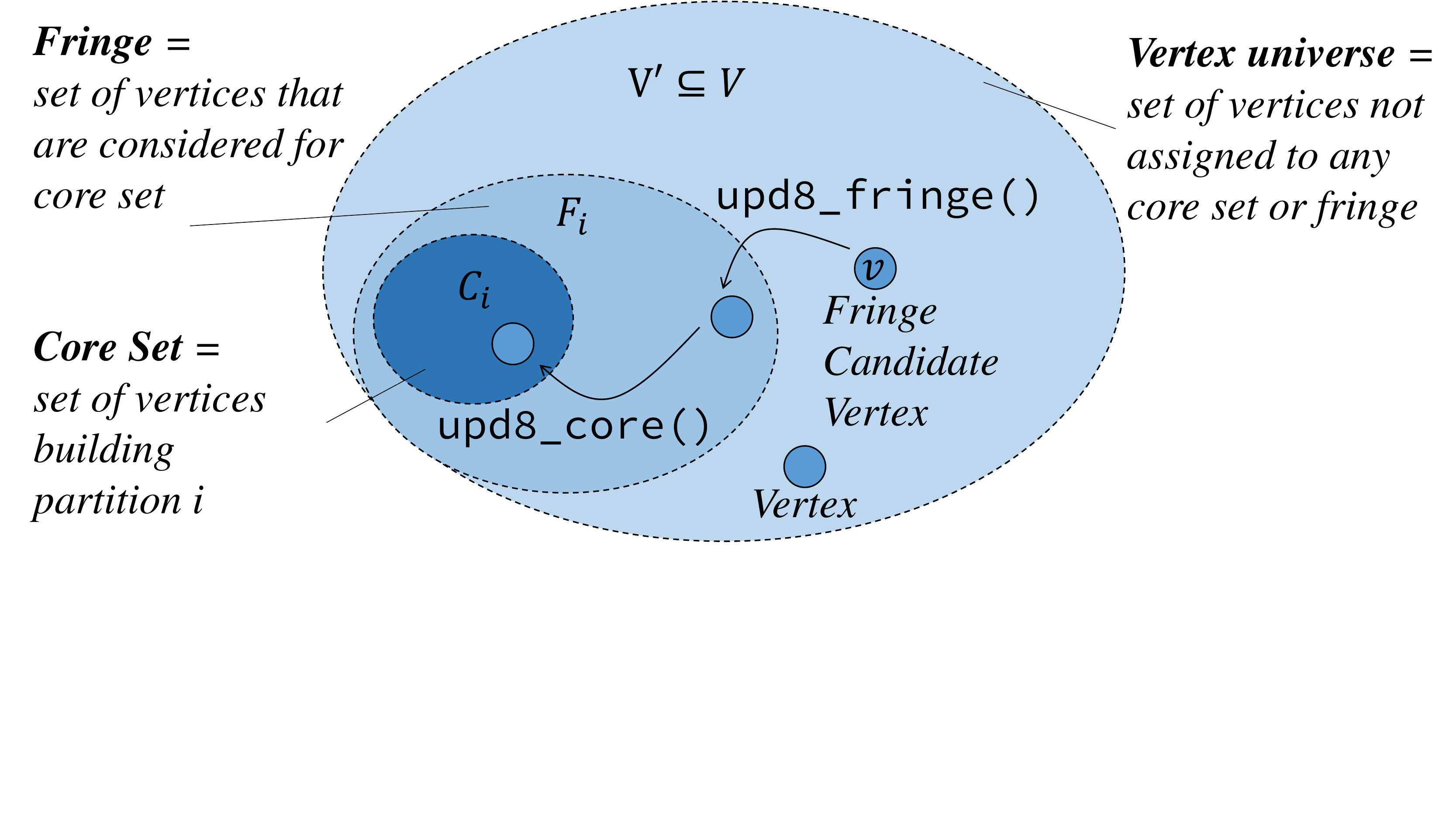}
	\vspace{-10pt}
	\caption{High-level idea of neighborhood expansion.}
	\label{fig:ideaNeighborhood}
	\vspace{-10pt}
\end{figure}

Figure~\ref{fig:ideaNeighborhood} sketches the general framework for growing the core set $C_i$ of partition $p_i \in P$. There are three overlapping sets: the vertex universe, the fringe, and the core set. 
The vertex universe $V' \subseteq V$ is the set of \textit{remaining vertices} that can potentially be added to the fringe $F_i$, i.e., $V' = V \setminus C_0... \setminus C_{i} \setminus F_i$. The fringe $F_i$ is the set of vertices that are currently considered for the core set. The core $C_i$ is the set of vertices that are assigned to partition $p_i \in P$. 
The three sets are non-overlapping, i.e., $V' \cap F_i = F_i \cap C_i = C_i \cap V' = \emptyset$. 

Initially, the core consists of seed vertices that are taken as a starting point for growing the partition. Based on these seed vertices, the fringe contains a subset of all neighboring vertices. In graph partitioning~\cite{Zhang:2017:GEP:3097983.3098033}, the fringe contains not a subset but all vertices that are in a neighborhood relation to one of the vertices in the core set. In the Figure~\ref{fig:ideaNeighborhood}, a fringe candidate vertex, say vertex $v$, is moved from the vertex universe to the fringe and then to the core set. In other words, any strategy based on neighborhood expansion must define the two functions \texttt{upd8\_fringe()} and \texttt{upd8\_core()}. 

As we develop a hypergraph partitioning algorithm based on the neighborhood expansion framework, we define the neighborhood relation and the three vertex sets accordingly. 
However, migrating the idea of neighborhood expansion from graph to hypergraph partitioning is challenging.
The number of neighbors of a specific vertex in a typical hypergraph is much larger than in a typical graph. The reason is that the number of neighbors is not only proportional to the number of incident hyperedges but also to the size of these hyperedges. 
For example, suppose you are writing a comment in the \texttt{/r/Python} subreddit. Suddenly, hundreds of thousands other authors in \texttt{/r/Python} are your direct neighbors. In other words, the neighborhood relation is \textit{group based} rather than \textit{bidirectional} which leads to massive neighborhoods.
The large number of neighbors changes the runtime behavior and efficiency of neighborhood expansion. For instance, in such a hypergraph, the fringe would suddenly contain a large fraction of the vertices in the hypergraph. But selecting a vertex from the fringe requires $\mathcal{O}(|V|)$ comparisons. 
This leads to high runtime overhead and does not scale to massive graphs.
HYPE alleviates this problem by reducing the search space significantly as described next.


\subsection{HYPE Algorithm}
\label{sec:alg:hype}

The HYPE algorithm grows the core set for partition $p_i \in P$ one vertex at a time. We load one vertex from the fringe to the core set and update the fringe with fresh vertices from the vertex universe. The decision of which vertex to include into the core and fringe sets is a critical design choice that has impact on the algorithm runtime and partitioning quality.

In this section, we explain the HYPE algorithm in detail, including a discussion of the design choices. In doing so, we first provide the basic algorithm in Section~\ref{sec:basicalgo} and then discuss optimizations of the algorithm in Section~\ref{sec:complexityreduction}. These optimizations tremendously reduce the algorithm runtime without compromising on partitioning quality.



\vspace{10pt}
\subsubsection{Basic Algorithm}
\label{sec:basicalgo}
The approach of growing a core set $C_i$ is repeated in a sequential manner for each partition $p_i$ for $i \in \{1,2,...,k\}$. It consists of a four step process. We initialize the computation with step 1., and iterate steps 2. and 3., until the algorithm terminates as defined in step 4.

\begin{enumerate}
	\item Initialize the core set.
	\item Move vertex from vertex universe into fringe.
	\item Move vertex from fringe into core set.
	\item Terminate the expansion.
\end{enumerate}

We now examine these steps in detail. A formal algorithmic description is given in Section~\ref{sec:alg:pseudocode}.

\begin{figure}
	\centering
	\subfloat[Partitioning quality.]{\label{fig:param:so:fringe_size1}   \includegraphics[width=0.24\textwidth]{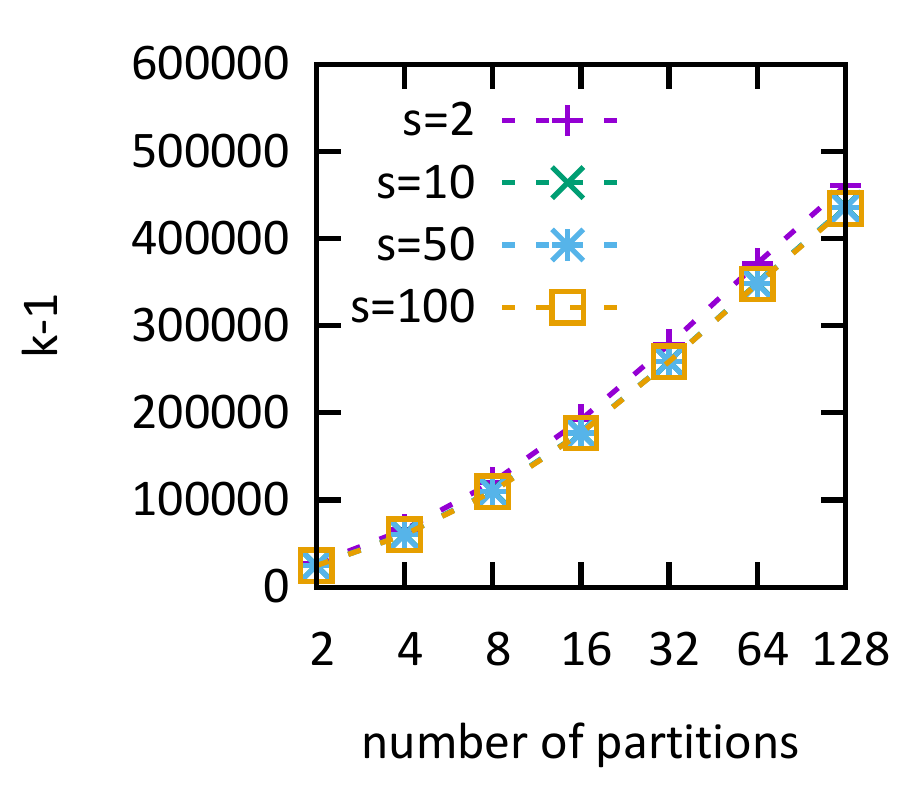}}
	\subfloat[Runtime.]{\label{fig:param:so:fringe_size2}   \includegraphics[width=0.24\textwidth]{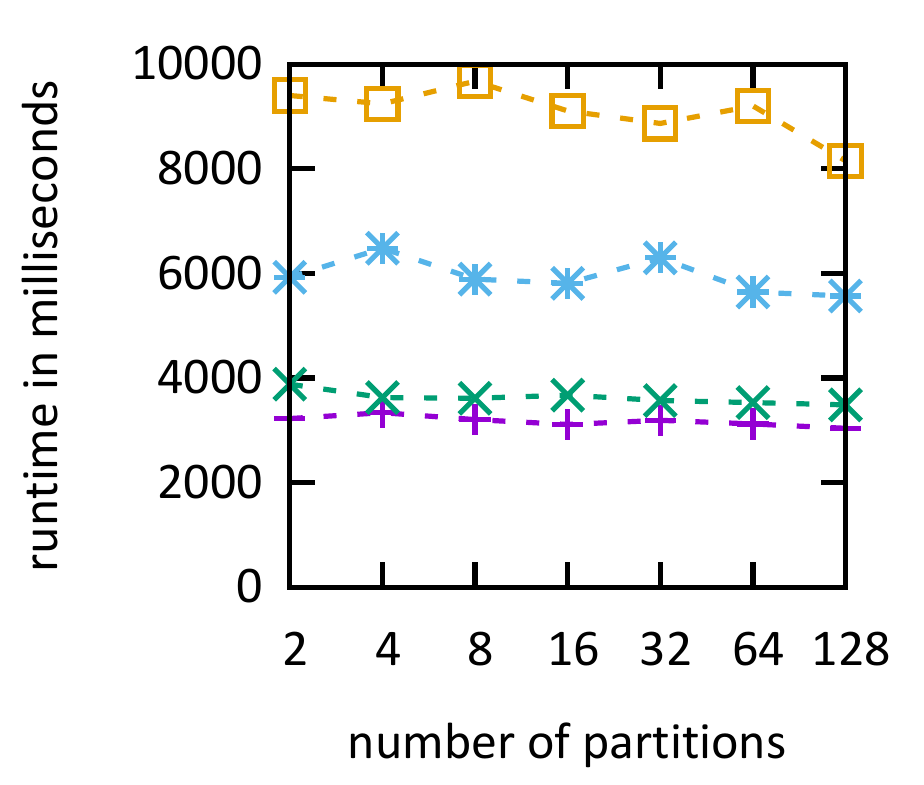}}
	\caption{ Limiting the fringe size $s$ to a small value (e.g. $s=10$) keeps partitioning quality intact while reducing runtime significantly (StackOverflow hypergraph).}
	\label{fig:param:so:fringe_size}
	\vspace{-10pt}
\end{figure}

\paragraph*{1. Initialize the core set}
The core set $C_i$ must initially contain at least one vertex in order to grow via the neighborhood expansion. In general, there are many different ways to initialize the core set. This problem is similar to the problem of initializing a cluster center for iterative clustering algorithms such as \texttt{K-Means} \cite{hartigan1979algorithm}. Here, the defacto standard is to select random points as cluster centers \cite{bradley1998refining, khan2004cluster}. In fact, a comparison of several initialization methods for \texttt{K-Means} shows that the random method leads to robust clustering quality \cite{pena1999empirical}. As the problem of selecting an initial ``seed'' vertex from which the core set grows is similar to this problem, we perform random initialization.
	
\paragraph*{2. Move vertex to fringe}
The function \texttt{upd8\_fringe()} determines which vertices move from the \textit{vertex universe} to the fringe $F_i$. The vertex universe consists of all vertices that are neither in the fringe $F_i$, nor in the core set $C_i$ of any previous execution of the algorithm for any partition $p_j \in P$ with $j<i$. 
	
The standard strategy of neighborhood expansion is to fill the fringe $F_i$ with all vertices that are neighbors to any core vertex, i.e., $F_i \gets N(C_i) \setminus C_0 \setminus ... \setminus C_i$. But for real-world hypergraphs, this quickly overloads the fringe with a large number of vertices from the vertex universe. To prevent this, we restrict the fringe to contain only $s$ vertices, i.e., $|F_i| \leq s$. In Figure~\ref{fig:param:so:fringe_size}, we validate experimentally that setting $s$ to a small constant value, i.e., $s=10$, keeps partitioning quality high but reduces runtime by a large factor. For brevity, we omit the discussion of similar results observed for other hypergraphs.

But how do we select the next vertex to be loaded into the fringe? 
Out of the vertex universe $V'$, we select a vertex to be included to the fringe using a scoring metric as described in the next paragraphs. The intuition behind this scoring metric is to find the vertex that preserves the highest \textit{locality} when assigned to the core set.  


To this end, we define \textit{locality} as the frequency that for a given vertex $v \in V'$, a neighboring vertex $v' \in N(v)$ resides on the same partition.
High locality leads to low cut sizes and good partitioning quality. 
To improve locality, our goal is to grow the core set into the smaller local communities and assign all vertices of these smaller communities to the same partition.
If a high proportion of neighbors of vertex $v \in V'$ is already assigned to the core set, assigning vertex $v$ to the core set as well will improve locality.

\begin{figure}	
	\includegraphics[width=0.45\textwidth, trim= 0pt 0pt 0pt 0pt, clip=true]{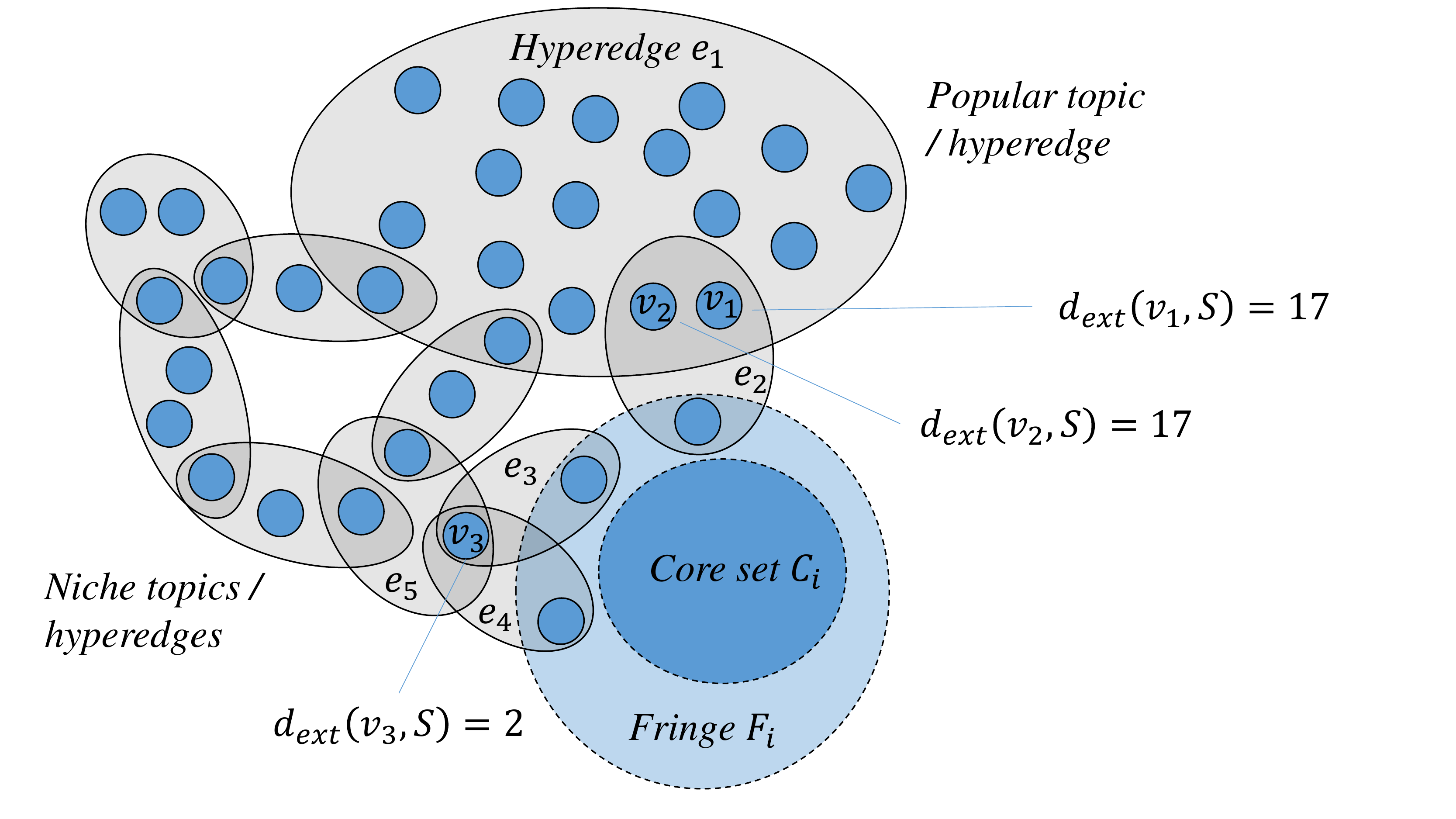}
	\vspace{-5pt}
	\caption{The external neighbors metric determines which vertex to move into the fringe.}
	\label{fig:externalNeighbors}
	\vspace{-10pt}
\end{figure}

In Figure~\ref{fig:externalNeighbors}, we see an example hypergraph that has the typical properties of real-world hypergraphs: the size of the hyperedges follows a power law distribution. To grow the fringe, there are three options: include vertex $v_1$, $v_2$, or $v_3$. Intuitively, we want to grow the fringe towards the local communities to preserve locality. We achieve this by selecting vertices based on the \textit{external neighbors} with respect to the fringe $F$.
In other words, we want to add vertices to the fringe that have a \textit{high} number of neighbors in the fringe or the core set, and a \textit{low} number of neighbors in the remaining vertex universe.
A vertex with low external neighbors score tends to have high locality in the fringe and the core set.
Formally, we write $d_{ext}(v,F_i)$ to denote the number of neighboring vertices of $v$ that are not already contained in the fringe as defined in Equation~\ref{eq:externalNeighbors}.

\begin{equation}
\label{eq:externalNeighbors}
d_{ext}(v,F_i) = |N(v) \setminus F_i|
\end{equation} 

We denote the vertices for which we calculate the external neighbors score as \textit{fringe candidate vertices}.
For each execution of \texttt{upd8\_fringe()}, we select $r$ fringe candidate vertices. The fringe $F_i$ contains up to $s$ vertices. Hence, after assigning one vertex to the core set, we take the top $s$ vertices out of the $s-1+r$ fringe candidate vertices as the new fringe. 

In Section~\ref{sec:complexityreduction}, we describe three optimizations on the \texttt{upd8\_fringe()} function that reduce the runtime while keeping the partitioning quality intact.

\paragraph*{3. Move vertex to core set}

Next, we describe the function \texttt{upd8\_core()} that moves a vertex from the fringe $F_i$ to the core set $C_i$. The function simply selects the vertex with smallest (cached) external neighbors score. This vertex $v$ is then moved to the core, i.e., $C_i \gets C_i \cup \{v\}$. This decision is final. Once assigned to the core $C_i$, vertex $v$ will never be assigned to any other core $C_j$ when considering a partition $j > i$. Hence, we remove the vertex from the remaining set of vertices in the vertex universe, i.e., $V' \gets V' \setminus \{v\}$. 
In case the fringe can not be filled with enough neighbors, we add a random vertex from the vertex universe to the fringe and proceed with the given algorithm.


\paragraph*{4. Terminate the expansion}

We terminate the algorithm as soon as the core set contains $\frac{|V|}{k}$ vertices. This leads to perfectly balanced partitions with respect to the number of vertices. Upon termination, we release the vertices in the fringe $F_i$ and store the vertices in the core set $C_i$ in a separate partitioning file for partition $i$. After this, we restart expansion for the next partition $i \gets i + 1$ or terminate if all vertices have been assigned to partitions. In Section~\ref{sec:balancingConsiderations}, we discuss other possible balancing schemes.

\renewcommand{\arraystretch}{1.8}
\begin{table*}
	{\footnotesize
		\centering
		\begin{tabular}{|l|l|l|l|l|l|}
			\hline
			Dataset & Vertices & Hyperedges & \#Vertices & \#Hyperedges & \#Edges \\ \hline\hline
			
			Github~\cite{Kunegis2013Konect}			& Users         & Projects   & 177,386   & 56,519    & 440,237\\
			\hline
			StackOverflow~\cite{Kunegis2013Konect} 	& Users         & Posts     & 641,876   & 545,196   & 1,301,942\\
			\hline
			Reddit   	& Subreddits    & Authors   & 430,156   & 21,169,586 & 179,686,265\\
			\hline
			Reddit-L      	& Comments    & Authors \& Subreddits   & 2,862,748,675   & 21,599,742 & 5,725,497,350 \\
			\hline
		\end{tabular}
		\caption{Real-world hypergraphs used in evaluations.}
		\label{tab:hypergraphs}
	}
	\vspace{-30pt}
\end{table*}

\vspace{10pt}
\subsubsection{Optimization of Fringe Updates}
\label{sec:complexityreduction}

When moving a vertex from the vertex universe to the fringe, we have to be careful in order to \emph{efficiently} select a good vertex. Calculating a score for all vertices in the vertex universe would be much too expensive. For example, suppose we add vertex $v_2$ to the fringe in the example in Figure~\ref{fig:externalNeighbors}. Suddenly, all vertices in the huge hyperedge $e_1$ could become fringe candidate vertices for which we would have to calculate the external neighbors score. Note that to calculate the external neighbors score, we must perform set operations that may touch an arbitrary large portion of the global hypergraph.

Our strategy to address this issue involves three steps: (a) select the best fringe candidate vertices from the vertex universe in an \emph{efficient} manner by traversing small hyperedges first, (b) reduce the number of fringe candidate vertices $r$ to $r=2$, and (c) reduce the computational overhead to calculate the score for a fringe candidate vertex by employing a scoring cache. These three optimizations help us to limit the overhead per decision of which vertex to include into the fringe. Next, we describe the optimizations in detail.

	
\paragraph{Maximize the chance to select $r$ \textbf{good} fringe candidate vertices}
	
First, we describe how we maximize the chance to select good fringe candidate vertices from the vertex universe.
The optimal vertex to add to the fringe has minimal external neighbors score, i.e., $argmin_{v \in V'} d_{ext}(v,F_i)$. Vertices that reside in large hyperedges (e.g. $e_1$ in Figure~\ref{fig:externalNeighbors}) have a high number of neighbors. Hence, it is unlikely that these vertices have a low external neighbors score with respect to the fringe $F$. For example, in Figure~\ref{fig:externalNeighbors}, vertices $v_1$ and $v_2$ have 18 neighbors, whereas vertex $v_3$ has only 4 neighbors. Thus, vertex $v_3$ has a much higher chance of being the vertex with the minimal external neighbors score. Based on this observation, we optimize the selection of fringe candidate vertices by ordering all hyperedges that are incident to the fringe $F_i$ with respect to their size and consider only vertices in the smallest hyperedge for inclusion into the fringe (e.g., hyperedges $e_4, e_3, e_2$ with $|e_4| \leq |e_3| \leq |e_2|$ results in the initial selection of hyperedge $e_4$). When we cannot retrieve $r$ vertices from the smallest hyperedge (because it does not contain enough vertices that are not already in $C$ or $F$), we proceed with the next larger hyperedge. 

\paragraph{Reduce the number of fringe candidate vertices to be selected}

\begin{figure}
	\centering
	\subfloat[Quality: (K-1) Metric.]{\label{fig:param:so:r1}   \includegraphics[width=0.24\textwidth]{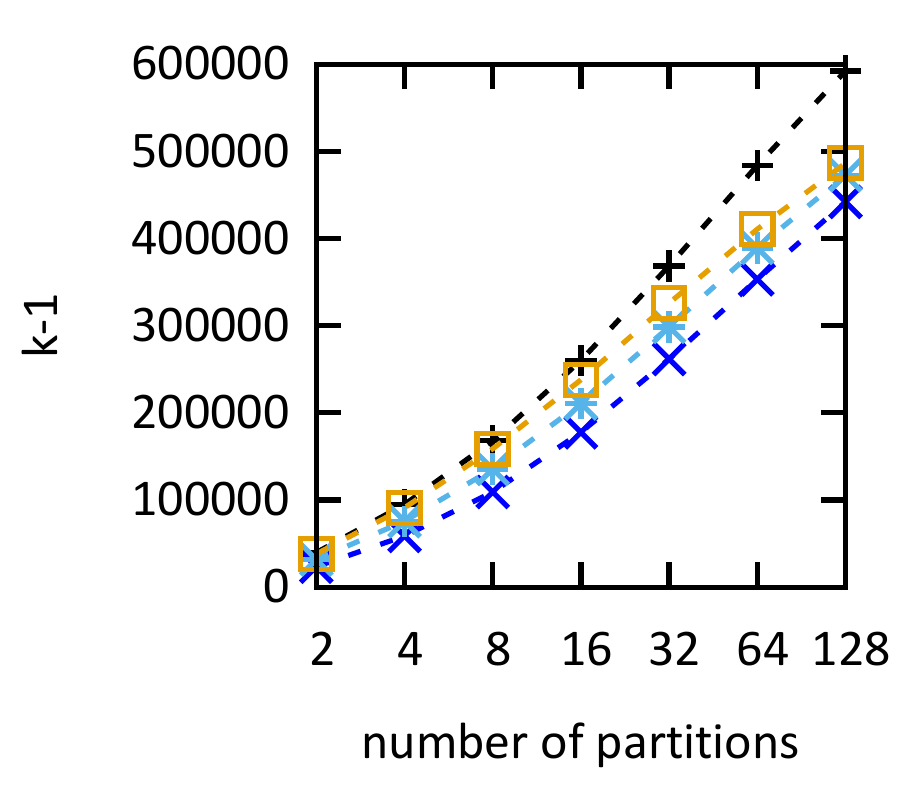}}
	\subfloat[Runtime.]{\label{fig:param:so:r2}   \includegraphics[width=0.24\textwidth]{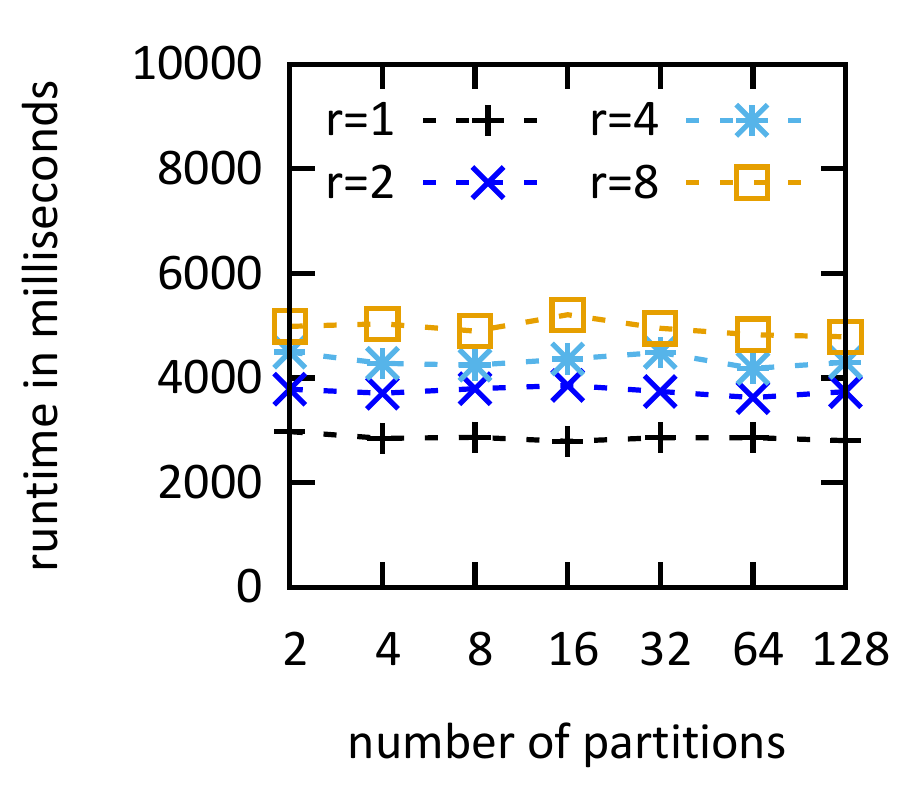}}
	\caption{Limiting the number of fringe candidate vertices $r$ to $r=2$ leads to the best partitioning quality (StackOverflow hypergraph).}
	\label{fig:param:so:r}
	\vspace{-10pt}
\end{figure}

Next, we limit the number of fringe candidate vertices to $r=2$ vertices.
From these $r$ vertices, we select the vertex with the smaller external neighbors score. 
The basic principle of selecting the best out of two random choices is known in literature as \textit{``the power of two random choices''} \cite{richa2001power} and has inspired our design. We experimentally validated that considering more than two options, i.e., $r > 2$,  does not significantly improve the decision quality, cf. Figure~\ref{fig:param:so:r}. Clearly, the lower the number of fringe candidate vertices $r$ is, the lower is the runtime of the algorithm. Interestingly, using two choices, i.e., $r=2$ leads to \textit{better} partitioning quality than all other settings of $r$. Apparently, a higher value for $r$ forces the algorithm to consider fringe candidate vertices from larger hyperedges which distracts the algorithm from the smaller hyperedges. 

\begin{figure}
\centering
\subfloat[Quality: (K-1) Metric.]{\label{fig:param:so:caching1}   \includegraphics[width=0.24\textwidth]{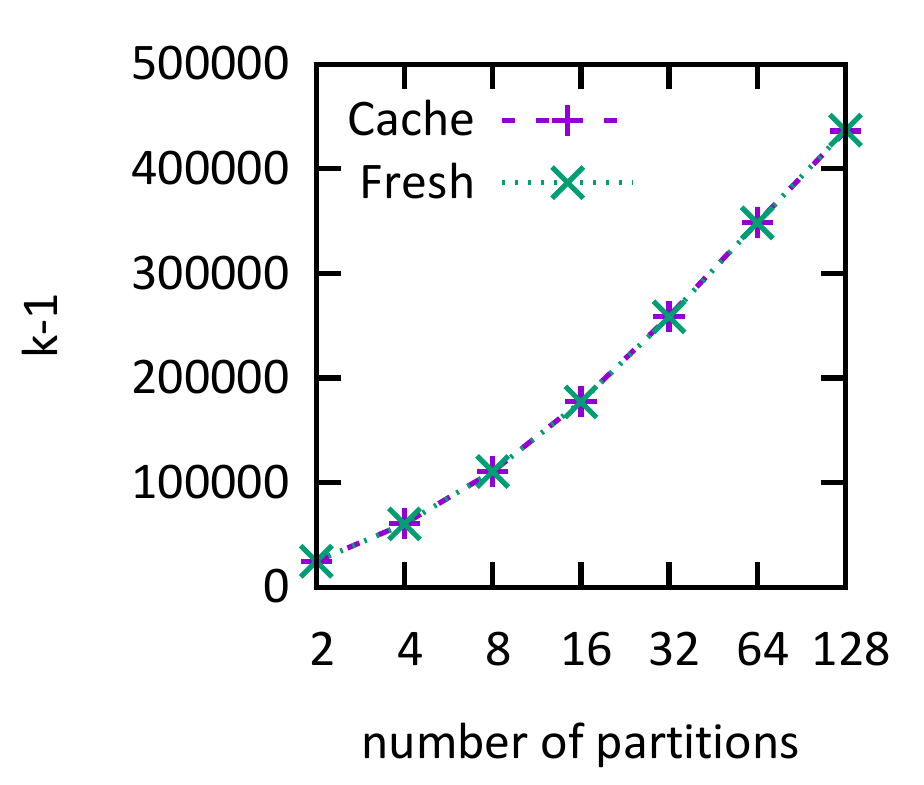}}
\subfloat[Runtime.]{\label{fig:param:so:caching2}   \includegraphics[width=0.24\textwidth]{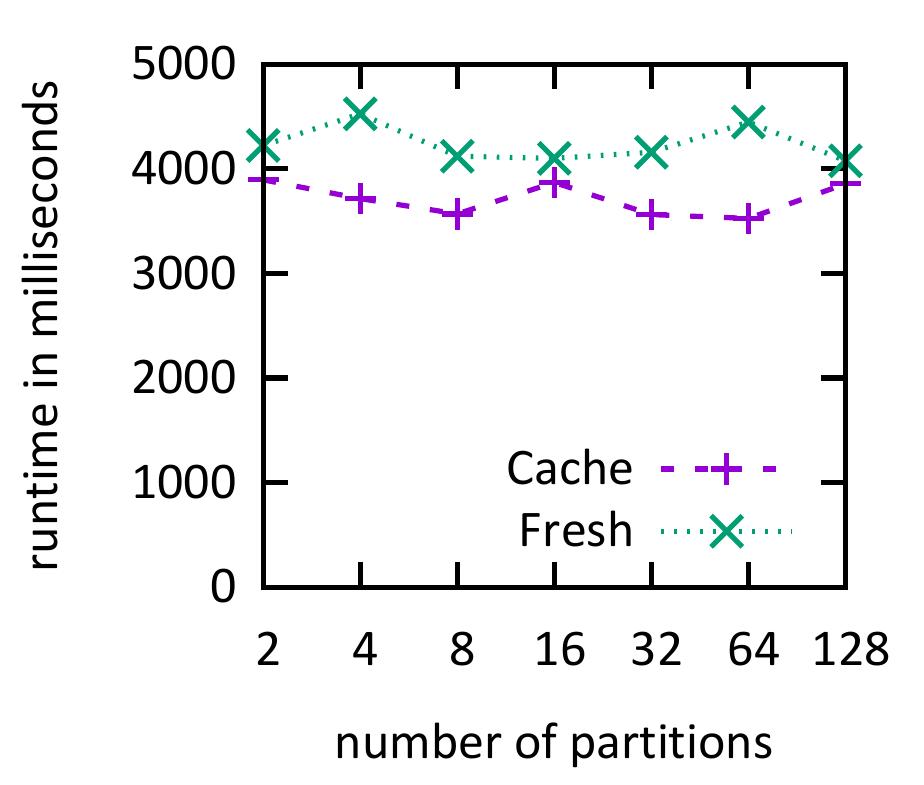}}
\caption{ The caching optimization for external neighbors score computation keeps partitioning quality intact while reducing runtime by up to 20\% on the Stackoverflow hypergraph.}
\label{fig:param:so:caching}
\vspace{-10pt}
\end{figure}

\paragraph{Reduce the overhead to compute an external neighbors score for fringe candidate vertices}

The external neighbors score requires calculation of the set intersection between two potentially large sets (see Equation~\ref{eq:externalNeighbors}). This calculation must be done for all fringe candidate vertices. 
To prevent recomputation, we use a caching mechanism. 
The score is calculated only when the vertex is accessed for the first time (lazy caching policy). While this means that the cached score may change when including more and more vertices into the fringe, our evaluation results show that partitioning quality stays the same when using caching (see Figure~\ref{fig:param:so:caching}). But the benefit of reducing score computations improves runtime by up to 20\%. 



\subsection{Balancing Considerations}
\label{sec:balancingConsiderations}

The default balancing objective of the HYPE algorithm leads to a balanced number of vertices on each partition. For $n$ vertices and $k$ partitions, the algorithm repeats the neighborhood expansion, one vertex at a time, until there are exactly $max = n/k$ vertices per partition. Vertex balancing is the standard method for distributed graph processing systems such as Pregel \cite{Malewicz2010Pregelb}---considering that the workload per partition is roughly proportional to the number of vertices per partition. Therefore, many practical algorithms such as the popular multilevel $k$-way hypergraph partitioning algorithm~\cite{Karypis:1999:MKW:309847.309954} focus on vertex balancing. 

However, some applications of hypergraph partitioning may benefit from balancing the sum of vertices \textit{and} hyperedges~\cite{Alistarh:2015:SMH:2969442.2969452}. More precisely, for $n$ vertices and $m$ hyperedges, the algorithm should partition the hypergraph in a way such that each partition is responsible for $\frac{n+m}{k}$ vertices or hyperedges. In the following, we discuss two ideas how HYPE can achieve this. First, we assign a weight $w(v)$ to each vertex $v$, i.e., the weight $w(v)=1+|E_v|$ with $E_v$ being the set of incident hyperedges of vertex $v$. Then, we repeat the neighborhood expansion algorithm by assigning vertices until each partition has $max = \frac{n+m}{k}$ total weight (or less). The rationale behind this method is the law of large numbers: it is not likely that a single vertex assignment will suddenly introduce a huge imbalance in relation to the already assigned vertices.
Second, to achieve perfect edge balancing, we can flip the hypergraph, i.e., viewing each original vertex as a hyperedge and each original hyperedge as a vertex. When balancing the number of vertices in the flipped graph, we actually balance the number of hyperedges in the original graph.
After termination of the algorithm, we flip the hypergraph back to the original representation.
We leave an investigation of other balancing constraints as future work.

\subsection{HYPE Pseudocode}

\label{sec:alg:pseudocode}


\begin{algorithm}
\small
	\caption{HYPE algorithm for hypergraph $G=(V,E)$.}
	\label{alg:HYPE}
	\begin{algorithmic}[1]
		\State $V' \gets V$
		\For{$i \in [0..k-1]$}
		\State $C_i \gets \{V'.random()\}$
		\State $V' \gets V' \setminus C_i$
		\State $F_i \gets \{\}$
		\State $c = <key,val>\{\}$ \Comment{clear cache}
		\While{$|C_i| < \frac{|V|}{k}$}
		\State \Call{upd8\_fringe}{$ $}		
		\State \Call{upd8\_core}{$ $}
		\EndWhile
		\EndFor	
	\end{algorithmic}
\end{algorithm}

Algorithm~\ref{alg:HYPE} lists the main loop. We repeat the following method for all partitions $p_i \in P$. After some housekeeping tasks such as filling the core set $C_i$ of partition $p_i$ with a random vertex (line 3), initializing the fringe (line 5), and clearing the $(key,value)$ cache (line 6), we repeat the main loop (lines 7-8) until the core set has exactly $\frac{|V|}{k}$ vertices. The loop body consists of the two functions \texttt{upd8\_fringe()} and \texttt{upd8\_core()} that are described next.

\begin{algorithm}
\small
	\caption{The function \texttt{upd8\_fringe()} updates the fringe $F_i$ with vertices from the vertex universe $V'$.}
	\label{alg:upd8fringe}
	\begin{algorithmic}[1]
		\Function{upd8\_fringe}{$ $}
		\State
		\State \#\textit{Determine $r$ fringe candidate vertices}
		\State $F_{cand} \gets \{ \}$ 
		\State $X \gets \{e \in E | C_i \cap e \neq \emptyset\}$
		\State $X' \gets [e_0, e_1, ... | e_j \in X \land |e_j|>|e_{j-1}|]$
		\For{$e \in X'$}
		\For{$v \in e \land v \not \in F_i \land \forall j \leq i: v \not \in C_j$} 
		\If{$|F_{cand}| < r$}
		\State $F_{cand} \gets F_{cand} \cup \{v\}$
		\Else
		\State \textbf{break} loop line 7
		\EndIf
		\EndFor
		\EndFor
		\State
		\State \#\textit{Update cache}
		\For{$v \in F_{cand} \land v \not \in c.keys()$}
		\State $c(v) \gets d_{ext}(v,F_i)$
		\EndFor  
		\State
		\State \#\textit{Update fringe}
		\State $F_i' \gets [v_0, v_1, ... | v_j \in F_i \cup F_{cand} \land c(v_{j-1})<c(v_j)]$
		\State $F_i \gets \{v | v \in F_i'.subsequence(0..s-1)\}$
		\If{$F_i = \emptyset$}
		\State $F_i \gets \{V'.random()\}$
		\EndIf
		\EndFunction
	\end{algorithmic}
\end{algorithm}

Algorithm~\ref{alg:upd8fringe} lists the \texttt{upd8\_fringe()} function. The function consists of three steps: determine the $r$ fringe candidate vertices (lines 3-10), update the cache (lines 12-14), and update the fringe (lines 16-20). 
The first step calculates the fringe candidate vertices $F_{cand}$ by first sorting the hyperedges that are incident to the core set $C_i$ by size (ascending) and then traversing these hyperedges for vertices that can still be added to the fringe (i.e., are not already assigned to the fringe or any core set).
The second step updates the cache with the current external neighbors score with respect to the current fringe $F_i$.
The third step sets the fringe to the set of top $s$ vertices with respect to the external neighbors score. If the fringe is still empty after these steps, we initialize it with a random vertex.

\begin{algorithm}
\small
	\caption{The function \texttt{upd8\_core()} updates the core $C_i$ with vertices from the fringe $F_i$.}
	\label{alg:upd8core}
	\begin{algorithmic}[1]
		\Function{upd8\_core}{$ $}
		\State $v \gets argmin_{v \in F_i} c(v)$
		\State $C_i \gets C_i \cup \{v\}$
		\State $F_i \gets F_i \setminus \{v\}$
		\State $V' \gets V' \setminus \{v\}$
		\EndFunction		
	\end{algorithmic}
\end{algorithm}

Algorithm~\ref{alg:upd8core} lists the \texttt{upd8\_core()} function. We load the vertex with the minimal cached external neighbors score into the core $C_i$ and remove this vertex from the fringe $F_i$ and the vertex universe $V'$.

\subsection{Complexity Analysis}
\label{sec:alg:complexity}
For the following analysis, we denote the number of vertices as $n=|V|$ and the number of hyperedges as $m=|E|$. Algorithm~\ref{alg:HYPE} repeats for $k$ partitions the procedure of moving $\frac{n}{k}$ vertices from the vertex universe to the fringe and from the fringe to the core. Next, we analyze the runtime for those procedures \texttt{upd8\_fringe()} and \texttt{upd8\_core()}. 

The function \texttt{upd8\_fringe()} in Algorithm~\ref{alg:upd8fringe} consists of three steps. First, it determines $r$ fringe candidate vertices from the vertex universe (lines 3-12). As we set $r$ to a small constant ($r=2$), this step is very fast in practice with the following caveat: The algorithm needs to sort the incident hyperedges with respect to the hyperedge size (line 6). The computational complexity of sorting all hyperedges is $O(m * log(m))$. It is sufficient to sort the hyperedges only once in the beginning of the HYPE algorithm. 
Recap that the algorithm fills the fringe with $r=2$ new candidate vertices. In the worst case, the fringe is incident to all hyperedges in the hypergraph. Therefore, selecting the fringe candidate vertices is in $O(m)$ to iterate over all hyperedges. \textbf{\emph{This is a pessimistic estimation---in practical cases it is sufficient to check the first few smallest hyperedges to find the $r=2$ candidates}}. 
Second, the algorithm updates the cache with fresh external degree scores for new candidates vertices (lines 14-16). It calculates the external degree score at most once for every candidate vertex (it is just read from cache if needed again later). As there are only $r=2$ fringe candidate vertices in each execution of \texttt{upd8\_fringe()}, the total number of external degree score calculations is limited to $2 * n$. The overhead of calculating the external degree of a vertex with respect to a set of $s$ fringe vertices is $O(s)$ (cf. Equation~\ref{eq:externalNeighbors}), but $s$ is a constant ($s=10$). Hence, the total computational complexity of updating the cache is $O(n)$. Third, the algorithm updates the fringe with vertices from the fringe candidates (lines 18-22). As both the fringe and the fringe candidates have constant sizes $r=2$ and $s=10$, the complexity of the third step is $O(1)$.

The function \texttt{upd8\_core()} in Algorithm~\ref{alg:upd8core} selects the vertex with minimal cached external neighbors score from the constant-sized fringe. Thus, the complexity is $O(1)$ including the housekeeping tasks in lines 3-5.

In total, the worst-case computational complexity of the HYPE algorithm is $O(m * log(m) + k * \frac{n}{k} * m + n  ) ) = O(m * log(m) + n * m)$. As highlighted above, in practice we observe that only a small, constant number of hyperedges is checked in order to find the $r$ candidate vertices, so that we observe a complexity of $O(m * log(m) + n)$.


\begin{figure*}
	\centering

	\subfloat[Quality:  (k-1) Metric.]{\label{fig:github:eval3}   \includegraphics[width=0.25\textwidth]{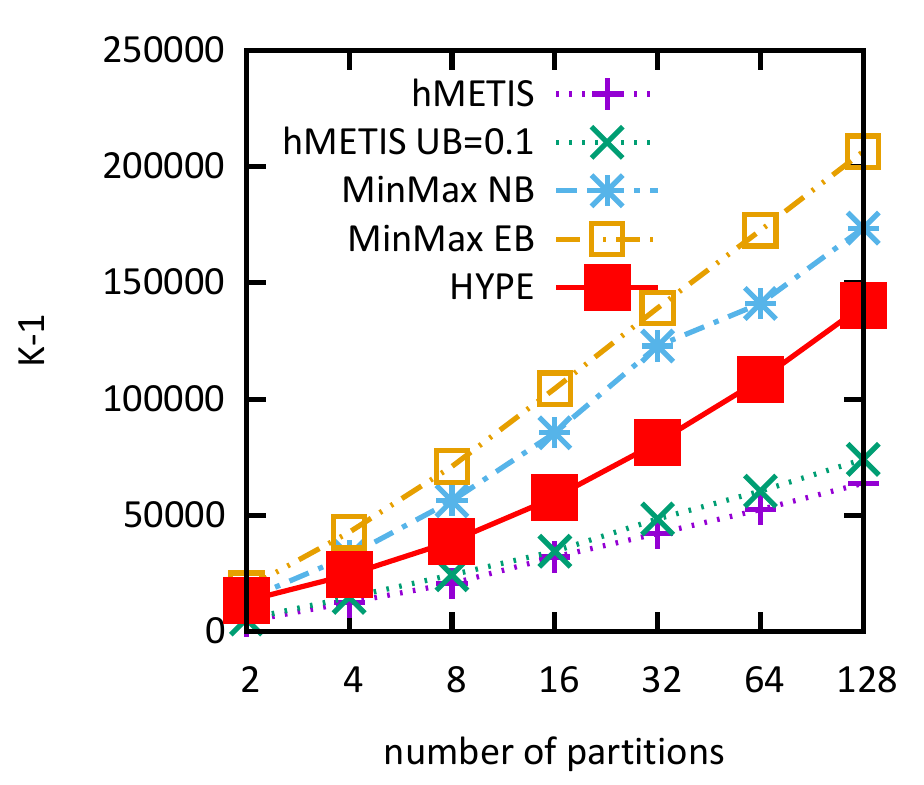}}
	\subfloat[Runtime.]{\label{fig:github:eval1}   \includegraphics[width=0.25\textwidth]{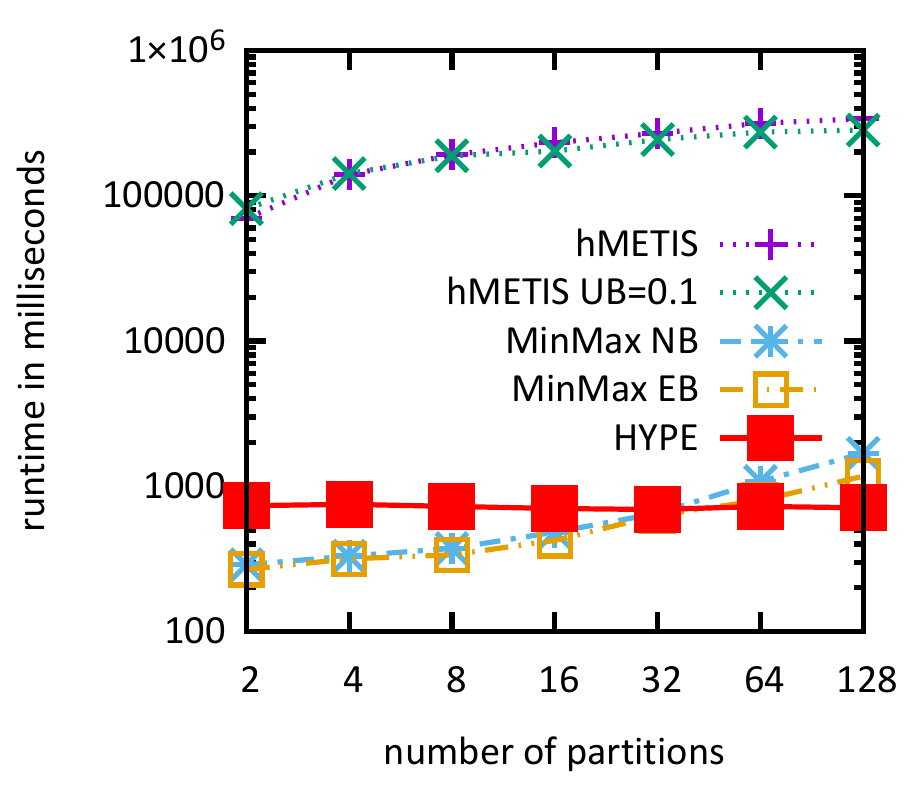}}
	\subfloat[Balancing.]{\label{fig:github:eval2}   \includegraphics[width=0.25\textwidth]{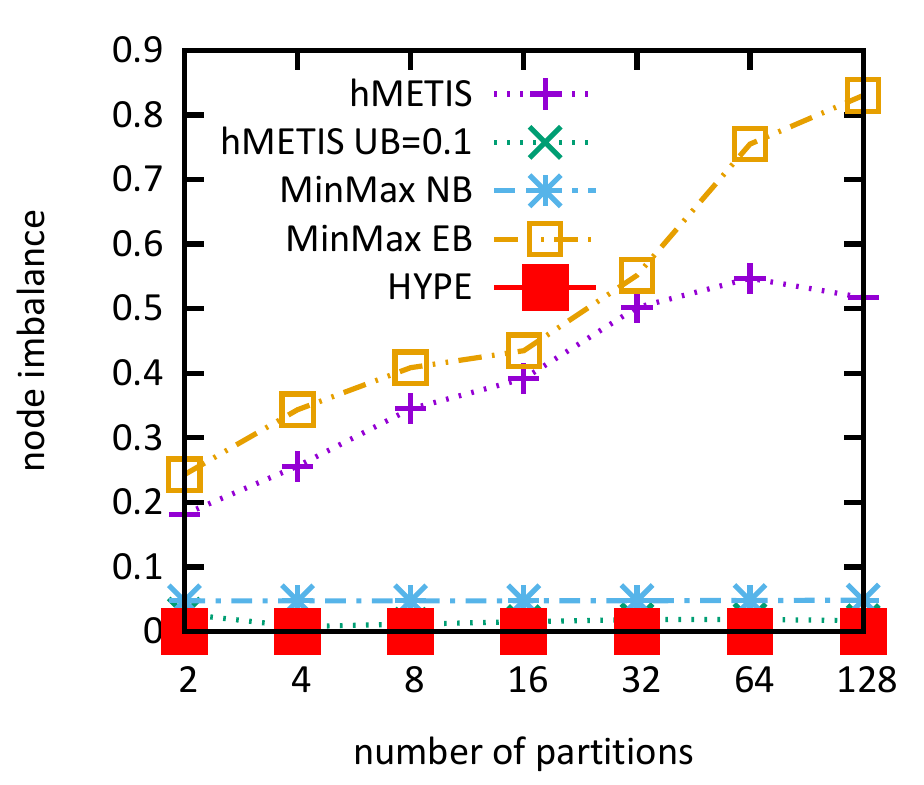}}
	
	\caption{Evaluations on the Github hypergraph (lower is better).}
	\label{fig:github:eval}
	\vspace{-20pt}
\end{figure*}

\section{Evaluations}
\label{sec:evaluations}

In this section, we evaluate the performance of HYPE on several real-world hypergraphs.

\paragraph*{Experimental Setup} All experiments were performed on a shared memory machine with 4 x Intel(R) Xeon(R) CPU E7-4850 v4 @ 2.10GHz (4 x 16 cores) with 1 TB RAM. The source code of our HYPE partitioner is written in C++.

\paragraph*{Hypergraph Data Sets} We perform the experiments on different real-world hypergraphs, i.e., Github~\cite{Kunegis2013Konect}, StackOverflow~\cite{Kunegis2013Konect}, 
Reddit and Reddit-L\footnote{\url{https://www.reddit.com/r/datasets/comments/3bxlg7/i_have_every_publicly_available_reddit_comment/}}, as listed in Table~\ref{tab:hypergraphs}. All of the hypergraphs show a power law distribution of vertex and hyperedge degrees. 
In addition to the number of vertices and hyperedges, we report the number of edges. An edge represents an assignment of a vertex to a hyperedge.

We highlight that for this paper, we crawled two large real-world hypergraphs from the Reddit dataset using the relations between authors, subreddits and comments. 

\paragraph*{Benchmarks} We choose our benchmarks for evaluating HYPE based on 3 categories.

Group (I) consists of a wide range of hierarchical partitioners such as hMetis~\cite{Karypis:1999:MKW:309847.309954}, Mondriaan~\cite{vastenhouw2005two}, Parkway~\cite{TRIFUNOVIC2008563}, PaToH~\cite{780863}, Zoltan~\cite{1639359}, and KaHyPar~\cite{hs2017sea}. 
As no partitioner in group (I) consistently outperforms the other partitioners in terms of partitioning quality, scalability and partitioning time, we decided for the well-established and widely used hypergraph partitioner hMETIS. Several recent papers show that hMETIS leads to competitive partitioning performance with respect to the $ (k-1)$ metric \cite{kabiljo2017social, yang2018hepart, Alistarh:2015:SMH:2969442.2969452}. 
Hence, we chose hMETIS as the representative partitioner from group (I) in this paper. We run hMETIS in two different settings: with and without enforced vertex balancing. 
Due to the high partitioning quality, hMETIS serves as the benchmark for partitioning quality on small to medium hypergraphs. We used hMETIS in version 2.0pre1 with the parameters \texttt{-ptype=rb -otype=soed -reconst}. To enforce vertex balancing, we set\footnote{We determined the UB parameter experimentally such that the measured imbalance is comparable to HYPE and MinMax. Schlag et al.~\cite{schlag2016k} provide an equation to set UB in order to enforce a specific imbalance constraint at a given k.} the parameter \texttt{UB=0.1}.


Group (II) consists of the recently proposed Social Hash Partitioner (SHP)~\cite{kabiljo2017social}. The authors released the raw source code of SHP. Yet, we could not reproduce their results as neither configuration files and parameters, nor scripts, execution instructions, or documentations were provided. 
However, in our evaluations we partitioned hypergraphs of similar size as SHP in a similar runtime, even though HYPE uses a purely sequential partitioning algorithm.

Group (III) comprises multiple streaming partitioning strategies proposed by Alistarh et al.~\cite{Alistarh:2015:SMH:2969442.2969452}. The greedy MinMax strategy constantly outperformed all other strategies according to their paper\footnote{The optimization goal of MinMax is to minimize the maximum number of hyperedges associated with a partition via the partition's vertices. However, both metrics MinMax and (k-1) are closely related: They measure the spread of hyperedges across partitions.}. 
Moreover, we designed a novel vertex-balanced variant of MinMax that outperforms the original approach with respect to the  (k-1) metric\footnote{We allowed a slack parameter of up to 100 vertices, cf. \cite{Alistarh:2015:SMH:2969442.2969452}.}. We denote this variant as \textit{MinMax NB} (\textbf{n}ode \textbf{b}alanced) in contrast to the standard \textit{MinMax EB} (\textbf{e}dge \textbf{b}alanced).

\begin{figure}[h]
	\centering
	
	\subfloat[Quality:  (k-1) Metric.]{\label{fig:so:eval3}   \includegraphics[width=0.25\textwidth]{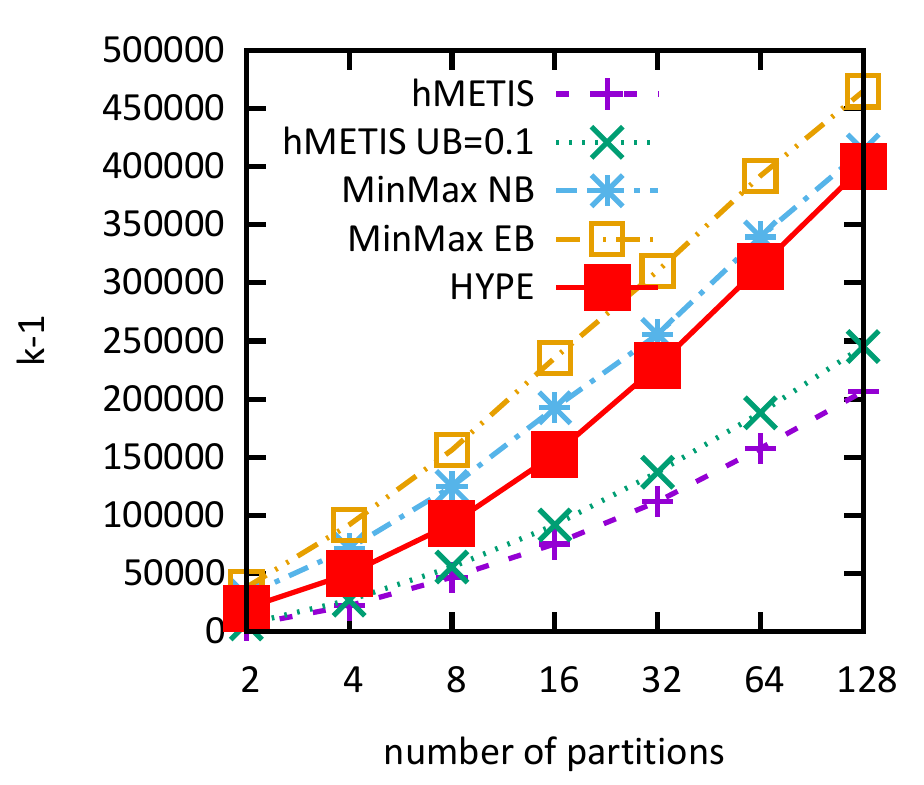}}
	\subfloat[Runtime.]{\label{fig:so:eval1}   \includegraphics[width=0.25\textwidth]{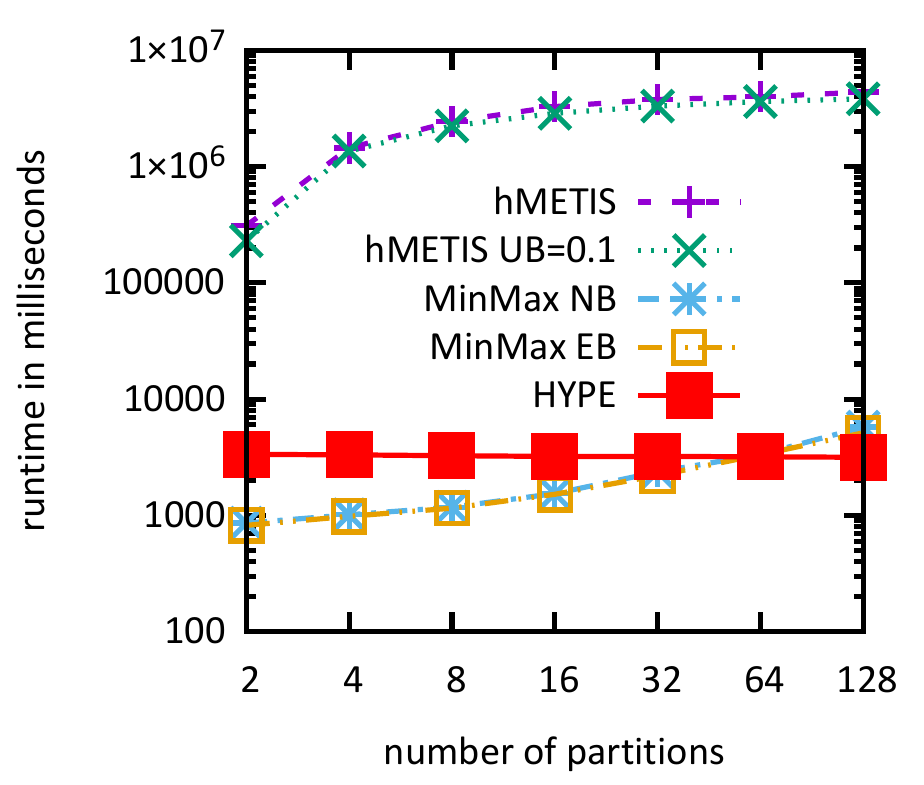}}
	
	\caption{Evaluations on the StackOverflow hypergraph (lower is better).}
	\label{fig:so:eval}
	\vspace{-16pt}
\end{figure}

\paragraph*{Experiments}

In all experiments, we capture the following metrics. (1) The (k-1) metric to evaluate the \emph{quality} of the hypergraph partitioning. This is the default metric for partitioning quality \cite{Karypis:1999:MKW:309847.309954, 1639359}. Other partitioning quality metrics such as the \textit{hyperedge-cut} and the \textit{sum of external degree} performed similar in our experiments\footnote{The close relationship between these metrics is described in literature \cite{kabiljo2017social}.}. (2) The runtime of the algorithm to partition the whole input hypergraph in order to evaluate the \emph{speed} of the partitioning algorithms.  (3) The vertex imbalance as a metric to capture the \emph{fairness} of the hypergraph partitioning. We compute vertex imbalance as the normalized deviation between the maximal and the minimal number of vertices assigned to any partition, i.e., $\frac{maxsize-minsize}{maxsize}$.

For each experiment, we increase the number of partitions from 2 up to 128 in exponential steps. 

\subsection{Performance Evaluations}

The performance evaluations show the benefits of HYPE when partitioning large hypergraphs. Its runtime is independent of the number of partitions, so that it is faster than streaming partitioning when the number of partitions is large. Further, the hierarchical partitioning algorithm hMETIS does not scale to very large hypergraphs, i.e., it cannot partition the Reddit hypergraph, and takes orders of magnitude longer for the smaller hypergraphs. We discuss the detailed results next.

\paragraph*{Github}
Figure~\ref{fig:github:eval3} shows the partitioning quality on the Github hypergraph. In the (k-1) metric, hMETIS performs best (e.g., 47 \% better than HYPE at k = 128). HYPE performs up to 45\% better than MinMax hyperedge-balanced, and up to 34~\% better than MinMax vertex-balanced. 

Partitioning runtime is depicted in Figure~\ref{fig:github:eval1}. hMETIS took 70 to 338 seconds to partition the Github hypergraph, which is orders of magnitude slower than MinMax and HYPE (up to 476 $\times$ slower than HYPE). The partitioning runtime of HYPE is independent of the number of partitions, as each partition is filled with vertices sequentially until it is full. Different from that, in MinMax, the partitioning runtime depends on the number of partitions, as MinMax works with a scoring function that computes for each vertex a score for each partition and then assigns the vertex to the partition where its score is best. Hence, for up to 32 partitions, HYPE has a higher runtime than MinMax, up to 2.7 $\times$ higher, whereas for 64 and 128 partitions, the runtime of HYPE is lower (up to 2.4 $\times$ lower).

In terms of balancing, cf. Figure~\ref{fig:github:eval2}, HYPE shows perfect vertex balancing, while the MinMax vertex-balanced has a slight imbalance of up to 5\%. Unsurprisingly, MinMax hyperedge-balanced has a poor vertex balancing. In hMETIS, when vertex balancing is turned on, the maximum imbalance was 3\%, whereas without that flag, vertex imbalance was tremedously higher (up to 55\% imbalance). The balancing results are similar for all other hypergraphs, so we will not discuss balancing in the following results any more.

\paragraph*{StackOverflow}
Figure~\ref{fig:so:eval3} shows the partitioning quality on the StackOverflow hypergraph.  In the  (k-1) metric, hMETIS performs best (e.g., 39 \% better than HYPE at k = 128). HYPE performs up to 47\% better than MinMax hyperedge-balanced, and up to 35\% better than MinMax vertex-balanced.

Partitioning runtime is depicted in Figure~\ref{fig:so:eval1}. hMETIS took between 206 and 4374 seconds (i.e., 73 minutes)  to partition the StackOverflow hypergraph, which is orders of magnitude slower than MinMax and HYPE (up to 1,384 $\times$ slower than HYPE). In numbers, HYPE takes 3 seconds to build 128 partitions, while hMETIS takes more than 1 hour! The comparison between HYPE and MinMax on StackOverflow leads to similar results as on the Github hypergraph: With up to 32 partitions, MinMax is faster (up to 4.1 $\times$ faster), but with 64 and 128 partitions, HYPE is faster (up to 1.8 $\times$ faster). Again, the reason for this is that the HYPE runtime is independent of the number of partitions.

\begin{figure}
	\centering
	
	\subfloat[Quality: (K-1) Metric.]{\label{fig:reddit:eval3}   \includegraphics[width=0.25\textwidth]{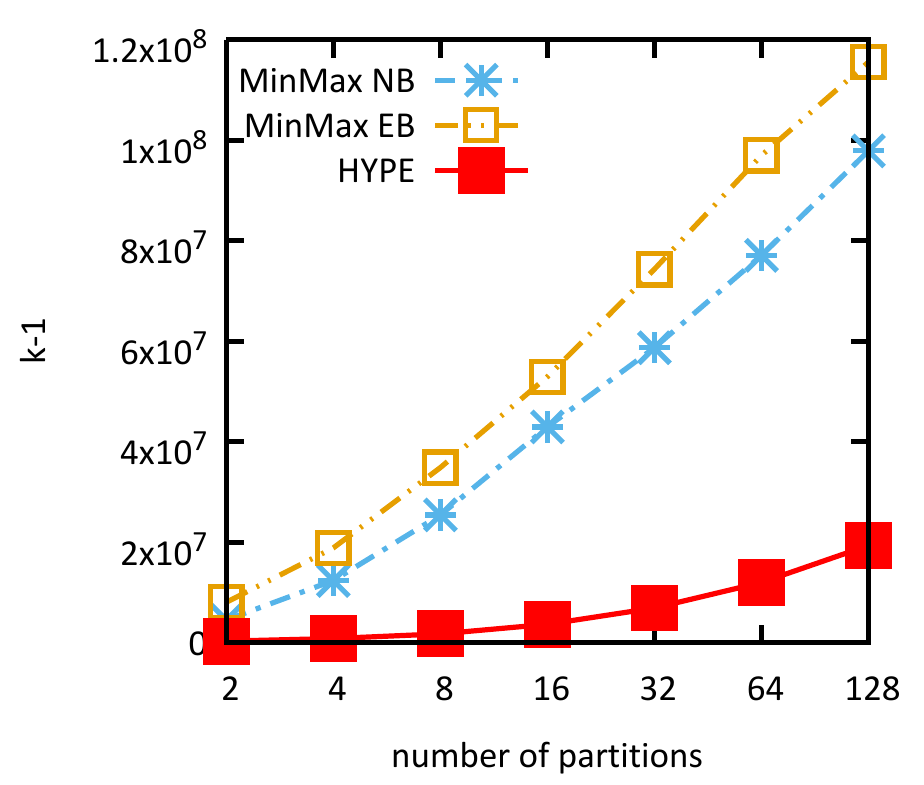}}
	\subfloat[Runtime.]{\label{fig:reddit:eval1}   \includegraphics[width=0.25\textwidth]{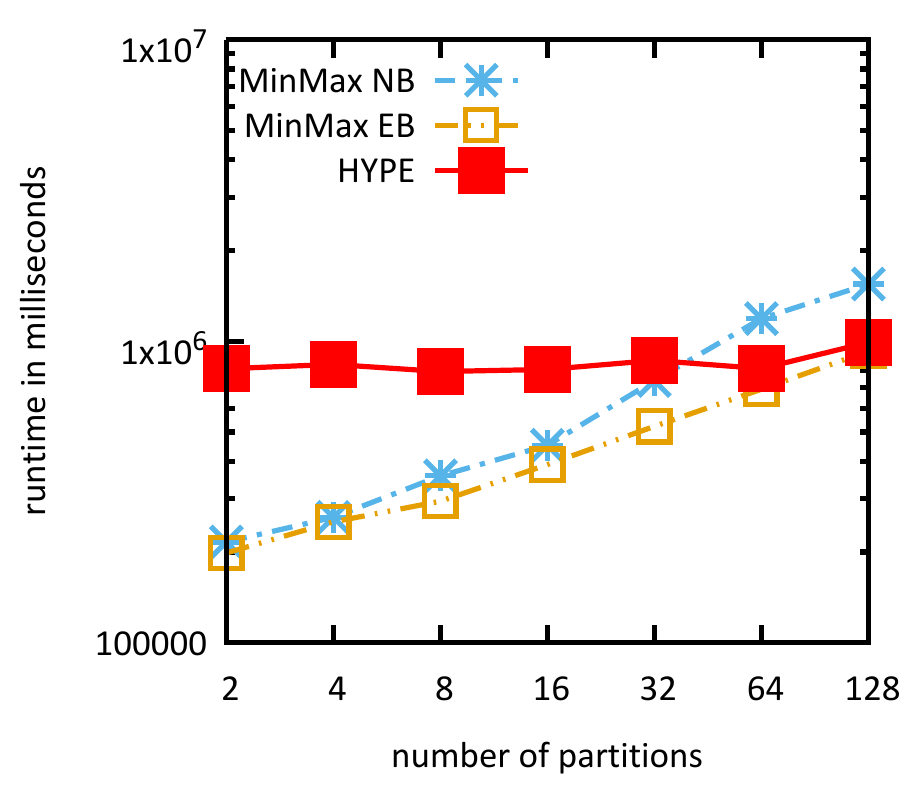}}
	
	\caption{Evaluations on the Reddit hypergraph (lower is better).}
	\label{fig:reddit:eval}
	\vspace{-12pt}
\end{figure}

%

\paragraph*{Reddit}
Figure~\ref{fig:reddit:eval3} shows the partitioning quality on the Reddit hypergraph. For this hypergraph, we could not produce results with the hMETIS partitioner, as it crashed or was running for days without returning a result. Consistent to the experiments reported in~\cite{kabiljo2017social}, many partitioners from group (I) are not able to partition such large hypergraphs. Hence, in the following, we only report results for MinMax and HYPE.

On the Reddit hypergraph, the advantage of exploiting local communities in HYPE pays off to the full extent: HYPE outperforms the streaming partitioner MinMax, that ignores the overall hypergraph structure, by orders of magnitude. For 2, 4 and 8 partitions, HYPE achieved an improvement of 95\% compared to MinMax hyperedge-balanced, and 93\% compared to MinMax vertex-balanced in the  (k-1) metric. Thus, HYPE leads to a partitioning quality that is up to \textbf{20 $\times$ better} than when using MinMax. For 16 partitions, HYPE performs 93\% and 91\% better, for 32 partitions 91\% and 88\%, for 64 partitions 88\% and 84\%, and for 128 partitions 83\% and 80\% better than MinMax hyperedge-balanced and MinMax vertex-balanced partitioners, respectively. 

Comparing the partitioning runtime of HYPE and MinMax in Figure~\ref{fig:reddit:eval1}, we see again that the runtime of HYPE is independent of the number of partitions, whereas MinMax has a higher runtime with growing number of partitions because of its scoring scheme. While at 2 partitions, MinMax is up to 4 $\times$ faster than HYPE, at 64 partitions HYPE becomes faster than MinMax, by up to 36\% at 128 partitions. As with the other hypergraphs, MinMax vertex-balanced is slightly slower than MinMax hyperedge-balanced as the hyperedge balance can change significantly after assigning a single vertex. This often forces assignment to a single partition (the least loaded) such that the partitions remain balanced. In such cases, the forced partitioning decisions for hyperedge-balanced partitioning can be performed very quickly. 

\begin{figure}
	\centering
	\subfloat[Quality:  (k-1) Metric.]{\label{fig:redditL:eval3}   \includegraphics[width=0.2\textwidth]{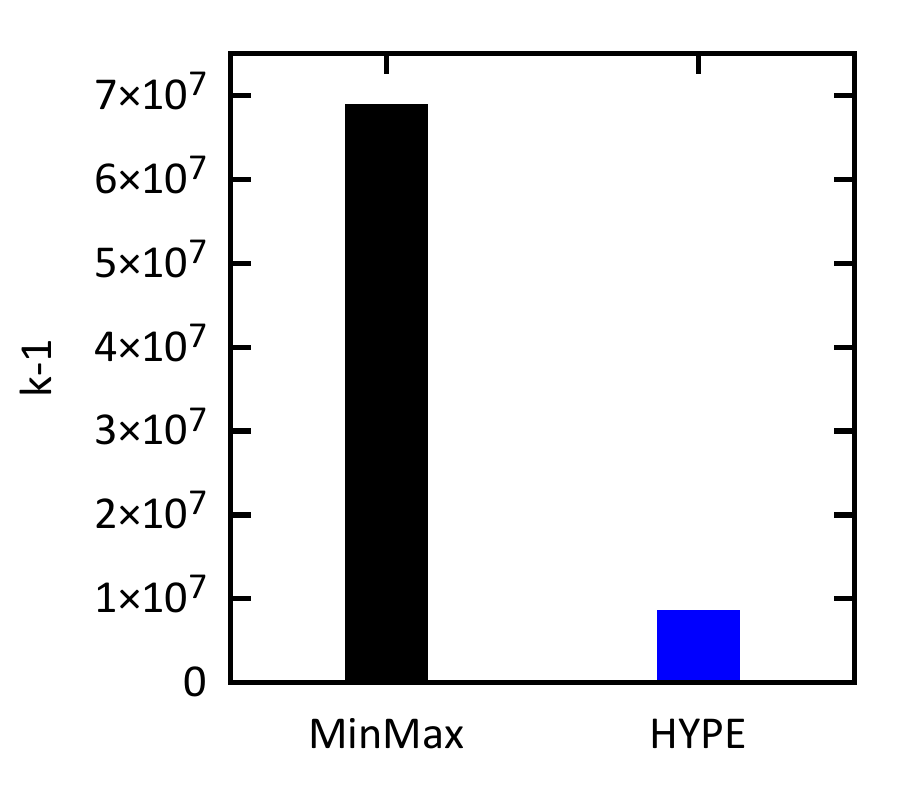}}
	\subfloat[Runtime.]{\label{fig:redditL:eval1}   \includegraphics[width=0.2\textwidth]{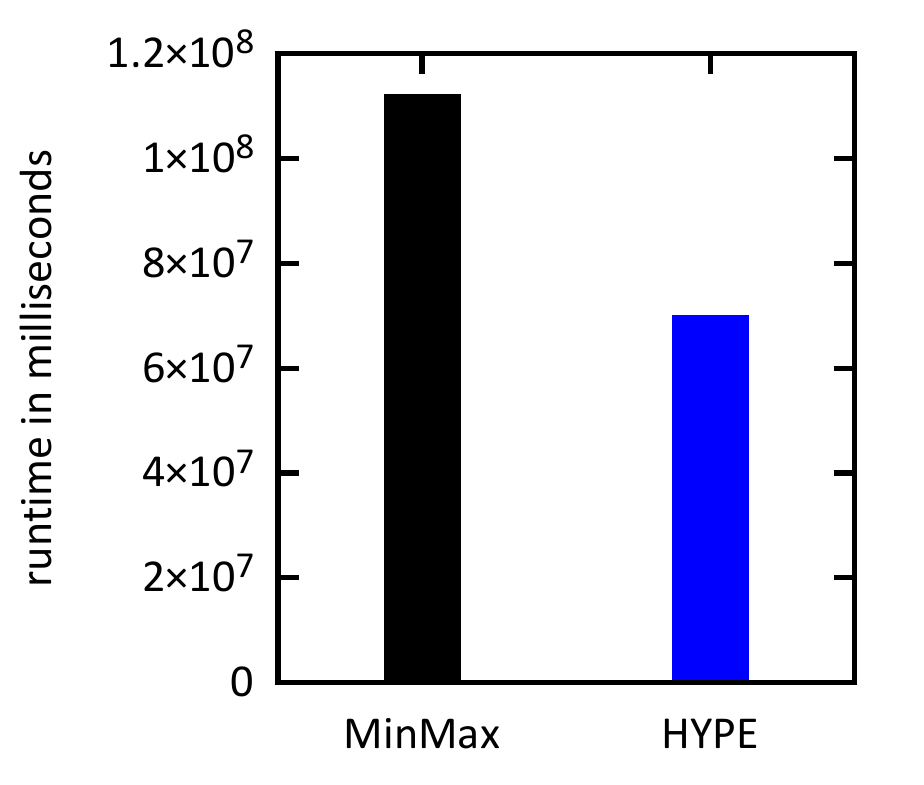}}
	
	\caption{Evaluations on the Reddit-L hypergraph (lower is better).}
	\label{fig:redditL:eval}
	\vspace{-12pt}
\end{figure}

\paragraph*{Reddit-L}
In Figure~\ref{fig:redditL:eval}, we compare partitioning quality and runtime of HYPE against MinMax on the large Reddit-L hypergraph with $k=128$ partitions. MinMax requires more than 31 hours to partition Reddit-L compared to the 19 hours of HYPE. Although being $39\%$ faster than MinMax, HYPE still outperforms MinMax in partitioning quality by $88\%$: MinMax has a  (k-1) score of 68,709,969 compared to HYPE's 8,357,200. Note that MinMax already belongs to the fastest high-quality partitioners. However, HYPE is able to outperform MinMax because its runtime does not depend on the number of partitions. 

\subsection{Discussion of the Results}

We conclude that HYPE shows very promising performance in hypergraph partitioning. First, it is able to partition very large hypergraphs, which cannot be partitioned by algorithms from group (I). Second, it consistently provides better partitioning quality than streaming MinMax.
On top of that, the HYPE algorithm is comparably easy to implement and to manage because all system parameters are fixed.

%
%
%

\section{Related Work}
\label{sec:relatedWork}


In the last decades, research on hypergraph partitioning was driven by the need to place transistors on chips in Very-large-scale integration (VLSI) design \cite{karypis1999multilevel}, as logic circuits can be modeled as large hypergraphs that are divided among chips. 
The most popular hypergraph partitioning algorithm from that area is hMETIS \cite{Karypis:1999:MKW:309847.309954} which is based on a multilevel contraction algorithm and produces good partitioning quality for medium-sized hypergraphs in the magnitude of up to 100,000 edges. 

However, multilevel partitioning algorithms do not scale to large hypergraphs, as shown in our evaluations. Parallel implementations of multilevel partitioning have been proposed \cite{1639359}, but the problem of high computational complexity and memory consumption remains. For instance, \textit{Zoltan}~\cite{1639359} is a parallel multilevel hypergraph partitioning algorithm. The evaluated graphs on Zoltan are relatively small---within a magnitude of up to 30 million edges or less---while using up to 64 parallel machines to process them. Other algorithms of that group are Mondriaan~\cite{vastenhouw2005two}, Parkway~\cite{TRIFUNOVIC2008563}, PaToH~\cite{780863}, and KaHyPar~\cite{hs2017sea}. For hypergraphs with hundreds of millions of edges, these algorithms are not practical as they take hours or even days to complete, if they terminate at all.


The bad scalability of multilevel partitioning algorithms led to the development of more scalable partitioners. \textit{Social Hash Partitioner (SHP)} achieves scalability to very large hypergraphs (up to 10 billion edges) by means of massive parallelization \cite{kabiljo2017social}. SHP performs random swaps of vertices between partitions and greedily chooses the best swaps. Random swaps fit well with the objective of parallelization and distribution in SHP, but may not be the most efficient heuristic. Investigating on the phenomenon of scalability versus efficiency~\cite{McSherry:2015:SBC:2831090.2831104}, we conceived the idea to devise an efficient hypergraph partitioning algorithm. 

Another approach to partition very large hypergraphs are \emph{streaming} algorithms. Streaming hypergraph partitioning takes one vertex at a time from a stream of vertices, and calculates a score for each possible placement of that vertex on each of the partitions. The vertex is then placed on the partition where its placement score is best, and cannot be removed any more. Alistarh et al. \cite{Alistarh:2015:SMH:2969442.2969452} proposed different heuristic scoring functions, where greedily assigning vertices to the partition with the largest overlap of incident hyperedges is considered best. There are two issues with the streaming approach. First, by only taking into account a single vertex at a time and placing it, information about the neighborhood of that vertex is not exploited although available in the hypergraph. Second, the complexity of the algorithm depends on the number of partitions, as the scoring function is computed for each vertex on each partition. For a large number of partitions, streaming partitioning becomes slow.
A closely related problem is \textit{balanced $k$-way graph partitioning} which faces similar challenges such as billion-scale graph data and the need for fast algorithms. Multilevel graph partitioners such as METIS \cite{karypis1998fast} and ParMETIS \cite{karypis1998parallel} do not scale very well. Spinner~\cite{7930049} is a highly scalable graph partitioner that, like SHP, performs iterative random permutations and greedy selection of the best permutation.
There is a large number of streaming graph partitioning algorithms, such as HDRF \cite{Petroni:2015:HSP:2806416.2806424}, H-load \cite{8263157}, and ADWISE \cite{Mayer2018ADWISE}. The ``neighborhood heuristic'' by Zhang et al. \cite{Zhang:2017:GEP:3097983.3098033} follows a completely different approach by exploiting the graph structure when performing partitioning decisions. The algorithm grows a core set by successively adding neighbors of the core set to a fringe set. However, the given heuristic can not be applied directly to hypergraph partitioning as the calculation of scores is way too expensive in hypergraphs (see Section~\ref{sec:algorithm}). 
While hypergraphs can be transformed into bipartite graphs, graph partitioning algorithms cannot be used to perform hypergraph partitioning. First, the bipartite graph representations contain one artificial vertex per hyperedge that destroys the vertex balancing requirement of hypergraph partitioning. Second, the (k-1) metric is ignored by graph partitioning algorithms. 

In recent years, several distributed hypergraph systems emerged that fueled the need for efficient massive hypergraph partitioning. These systems are inspired from the area of distributed graph processing systems and apply the vertex-centric programming model from graph processing to hypergraph processing. For instance, HyperX~\cite{7373388} allows applications to specify vertex and hyperedge programs which are then executed iteratively by the system. Also, Mesh~\cite{heintz2016mesh} builds upon the popular GraphX system~\cite{gonzalez2014graphx} and shows promising performance. These systems show significant reduction of processing latency with improved partitioning quality.
\section{Conclusions}
\label{sec:conclusion}


In this paper, we propose HYPE, an effective and efficient partitioner for real-world hypergraphs. 
The partitioner grows $k$ core sets in a sequential manner using a neighborhood expansion algorithm with several optimizations to reduce the search space.
Due to the simplicity of the design and focus on the hypergraph structure, HYPE is able to partition the large Reddit hypergraph with billions of edges in less than a day. This is the partitioning of one of the largest real-world hypergraph reported in literature. HYPE not only improves partitioning quality by up to $95\%$ compared to streaming hypergraph partitioning, but \textit{reduces} runtime as well by $39\%$.

A promising line of future research on HYPE is to explore how to grow the $k$ core sets in parallel. In this scenario, several core sets \textit{``compete''} for inclusion of attractive vertices, so the crucial questions are how to minimize the number of ``collisions'' and how to deal with collisions when they happen.  

\balance

\bibliographystyle{IEEEtran}
\bibliography{bib}

\end{document}